\documentclass[a4paper,12pt]{article}
\usepackage{mathptmx}
\usepackage{epsfig}
\usepackage{float}
\usepackage{amssymb}
\usepackage{natbib}
\usepackage{latexsym}
\usepackage{amsfonts}
\newcommand{\beq}{\begin{equation}}
\newcommand{\eeq}{\end{equation}}
\newcommand{\beqa}{\begin{eqnarray}}
\newcommand{\eeqa}{\end{eqnarray}}

\begin{document}\large
\title{\bf Hadron Generator and Atmospheric Seasonal Variation Influence on
Cosmic Ray Ionization computed with CORSIKA Code }

\author{A. Mishev$^{1}$ and P.I.Y. Velinov$^{2}$\\ \\
{\it $^{1}$ \small Institute for Nuclear Research and Nuclear Energy-Bulgarian Academy of Sciences,} \\
{\it \small Tsarigradsko Chaussee 72, 1784 Sofia, Bulgaria}  \\
{\it $^{2}$ \small Institute for Space Research and Technology, Bulgarian Academy of Sciences}\\
{\it \small Bl. 1 Acad. G. Bonchev str. 1113 Sofia, Bulgaria}
\\ \\
}
\normalsize

\maketitle
\begin{abstract}
Recently an essential progress in development of physical models for cosmic
ray induced ionization in the atmosphere is achieved. Basically, the models are full target, i.e. based on Monte Carlo simulation of an electromagnetic-muon-nucleon cascade in the atmosphere. In general, the contribution of proton nuclei in those models is highlighted, i.e. primary cosmic ray $\alpha$-particles and heavy nuclei are neglected or scaled to protons. The development of cosmic ray induced atmospheric cascade is sensitive to the energy and mass of the primary cosmic ray particle. The largest uncertainties in Monte Carlo simulations of a cascade in the Earth atmosphere are due to assumed hadron interaction models, the so-called hadron generators. In the work presented here we compare the ionization yield functions $Y$ for primary cosmic ray nuclei, such as protons, $\alpha$-particles, Oxygen and Iron nuclei, assuming different  hadron interaction models. The computations are fulfilled with the CORSIKA 6.9 code using GHEISHA 2002, FLUKA 2011, UrQMD hadron generators for energy below 80 GeV/nucleon and QGSJET II for energy above 80 GeV/nucleon. The observed difference between hadron generators is widely discussed. The influence of different atmospheric parametrizations, namely US standard atmosphere, US standard atmosphere winter and summer profiles on ion production rate is studied. Assuming realistic primary cosmic ray mass composition, the ion production rate is obtained at several rigidity cut-offs - from 1 GV (high latitudes) to 15 GV (equatorial latitudes) using various hadron generators. The computations are compared with experimental data. A conclusion concerning the consistency of the hadron generators is stated.
\end{abstract}


\label{cor}{\small Corresponding author:A.Mishev, INRNE-BAS, Tel:
(+359) 29746310; e-mail: mishev@inrne.bas.bg; also at {\it Sodankyl\"a Geophysical Observatory (Oulu unit), University of Oulu, Finland \\}
 \normalsize

\section{Introduction}

The Earth is constantly hit by elementary particles
and atomic nuclei of very large energies distributed in a wide energy range:
cosmic rays (CR)s. The primary CR flux variate from $10^4$
$m^{-2}$ $s^{-1}$ at energies $ \approx 10^{9}$ eV to $10^{-2}$ $km^{-2}$
$yr^{-1}$ at energies $\approx 10^{20}$ eV. The cosmic ray intensity is
approximatively expressed with (1), where E is the total particle
energy per nucleon in GeV and $\alpha = -2.7$ is the spectral index.
The majority of these particles are protons \citep{Nak10}. 

\begin{equation}
I_{n}(E)\varpropto1.8(\frac{E}{GeV})^{\alpha} nucleons.  cm^{-2}.
s^{-1}. sr^{-1}
\end{equation}

The abundance of primary CR is approximately independent of energy, at least
over the dominant energy range of 10 MeV/nucleon through several
GeV/nucleon. By mass about 79 $\%$ of nucleons in cosmic rays are
free protons, and about 80 $\%$ of the remaining nucleons are
bound in helium.

Most of the primary CR particles are of extra-solar origin known as galactic cosmic rays (GCR).
It is regarded that they mostly originate from the
Galaxy. The part below the "knee" comes from Galactic supernovas,
particles accelerated by the shocks in the supernova
remnants (SNR) see \citet{Kry77, Bla78, Ber09} and references therein. Those CR particles are always present in the vicinity of Earth. Their intensity is affected by solar activity, following the 11-year solar cycle and responding
 to long and short time scale solar-wind variations. Occasionally some solar flares and eruptive events, such as coronal mass ejections (CMEs) can accelerate protons and other ions to high energies
 \citet{Cli04, Aschwanden12} and references therein. Such solar energetic particles (SEPs) enter the atmosphere sporadically, with a greater probability during periods of high solar activity e.g. \citep{Shea90}.
Similar to GCR cascades described above, they could lead to an increase of the intensities recorded by neutron monitors on the surface of the Earth, known as ground level enhancements (GLEs).

The transport of CR particles is affected by the Earth's magnetosphere, which prevents penetration
 of charged particles, i.e. it provides a shielding effect, most effective near the geomagnetic equator. It is 
quantified by the effective vertical rigidity cut-off  $R_{C}$ defined as particle's momentum over charge.
In the work presented here we consider the effective vertical rigidity cut-off \citep{Cook91}, which varies with the geographical location.

The CR particles are the main contributor the the ionization in the middle atmosphere and troposphere \citep{Vel74, Dor04, Baz08, Uso09}. Primary CR particles initiate nuclear-electromagnetic-muon cascade resulting in an ionization of the ambient air. In this cascade only a fraction of the energy of the primary CR is transfer to high energy secondary particles reaching the ground. Most of the primary energy is released in the atmosphere by ionization and excitation of the molecules of air \citep{Baz08, Uso09}.

The ion pair production is related to various atmospheric processes \citep{Van03, Jag06, Baz08, Dor09}, influence on electric circuit and on chemistry compositions, aerosols etc. in the middle atmosphere, specifically during major solar proton events \citep{Vit96, Dam08, Jack08, Vel08, Cal11, Jack11, Mironova12, Kil13, Ton14, Tas14}. While the atmospheric effect of cosmic ray of galactic or solar origin is highly debated, the role of cosmic ray induced ionization is apparent \citep{Usos06, Baz08, MishevSta08, Dor09, Mishev10}. At present an essential progress in development of physical models for cosmic ray induced ionization in the atmosphere is achieved \citep{Usos06, Mis07a, Vel07, Baz08, Uso09, Vel09, Mishev12}. The estimation of cosmic ray induced ionization is possible on the basis of semi-empirical models \citep{Bri70}, simplified analytical models \citep{Bri05} or on a Monte Carlo simulation of the atmospheric cascade \citep{Des05, Uso04, Usos06, Vel09}. The analytical models are constrained to a given atmospheric region and/or cascade component or primary particles \citep{Vel12, Ase13}. 

The largest uncertainties in numerical simulations of an atmospheric cascade are due to the assumed models for hadron interactions i.e. the so-called hadron generators. The stochastic nature of the individual particle production leads to large shower-to-shower fluctuations, which depend on the particle mass number. The probability for interaction of a primary CR particle depends only on its traversed amount of matter (atmospheric air). The atmospheric depth associated with a given height above sea level plays a key role in cascade simulation \citep{Risse04, Engel11}. In addition,  the development of the cascade process in the atmosphere depends on the properties of the medium \citep{Ber00}. As it was recently demonstrated, the seasonal variations of the atmospheric profiles assumed in CORSIKA code seems to be rather large and they play an important role on cascade development simulation \citep{Kil04, Kil06}.

Moreover, the contribution of proton nuclei in a recent studies of cosmic ray induced ionization is highlighted \citep{Ond08, Mis10, Mis11, Cal11, Uso11}. Basically, the contribution of CR nuclei to the atmospheric ionization is neglected or scaled to protons \citep{Des05, Usos06, Mis10, Uso11}. In this connection, the influence of assumed low energy hadron interaction models and atmosphere seasonal variations in CORSIKA (COsmic Ray SImulations for KAscade) code on the energy deposit, respectively ionization is of a big interest. In the paper presented here, we study the effect of assumed hadron generators on computations of cosmic ray induced ionization in the atmosphere, specifically the contribution of heavy nuclei, as well as the influence of different atmospheric parametrizations assumed in CORSIKA code, namely US standard atmosphere, US standard atmosphere winter and summer profiles (for details see Appendix D in \citet{Hec97} and references therein).

\section{Numerical full-target model for cosmic ray induced ionization}
As was mentioned above, the Monte Carlo simulation of the atmospheric cascade allows to obtain the longitudinal cascade evolution in the atmosphere and the energy deposit by the different  cascade components from ground level till the middle atmosphere. Subsequently it is a matter of formalism to derive the ion rate production. The full target models apply the formalism of ionization yield function $Y$ similarly to Oulu model  \citep{Usos06}:

\begin{equation}
    Y(x,E) = \frac{\triangle E(x,E)\Omega}{\triangle x E_{ion}}
    \label{simp_eqn2}
   \end{equation}

where $\triangle E$ is the deposited energy in a atmospheric
layer $\triangle x$, $\Omega$ is the geometry factor - a solid
angle and $E_{ion}$ = 35 eV is the energy necessary for creation of an ion pair in air
\citep{Vel74, Por76}. We express $x$, during the simulations  in
$g/cm^{2}$, which is a residual atmospheric depth i.e. the amount
of matter (air) overburden above a given altitude in the
atmosphere. This is naturally related to the development of the
cascade. Subsequently the mass overburden is transformed as
altitude above sea level (a.s.l.) in [km].

The atmospheric ionization is obtained on the basis of equation
(3) following the procedure \citep{Mis07a, Vel07, Vel09}

\begin{equation}
    Q(x,\lambda_{m}) =  \sum_{i} \int_{E}^{\infty}D_{i}(E)Y_{i}(E,x)\rho(x)dE
        \label{simp_eqn2}
   \end{equation}

where $D_{i}(E)$ is the differential cosmic ray spectrum for a given
component $i$: protons p, Helium ($\alpha$-particles), Light nuclei L (3 $\le$ Z $\le$ 5), Medium nuclei M (6 $\le$ Z $\le$ 9), Heavy nuclei H (Z $\ge$ 10) and Very Heavy nuclei VH (Z $\ge$ 20) in the composition of primary cosmic rays (Z is the atomic number), $Y_{i}$ is the ionization yield function defined according equation (2) for various $i$, $\rho$ is the atmospheric density, $\lambda_{m}$ is the geomagnetic latitude, $E$ is the
initial energy of the incoming primary nuclei on the top of the
atmosphere. The geomagnetic latitude $\lambda_{m}$ is related to the
rigidity, therefore is related to the integration (integration above
$E$).

The ionization yield function Y depends only of assumed physical
models for cascade processes in the atmosphere, the atmospheric
model and the type of the primary particle. The major contribution of cosmic ray
induced ionization in the atmosphere is due to particles having
energy till 1 TeV, taking into account the steep spectrum of primary
cosmic rays (1). This is the reason to deal in this paper with
primary protons with maximal energy of 1 TeV.

The evolution of atmospheric cascade is performed with the CORSIKA 6.990 code \citep{Hec97}.  The assumed hadron generators are FLUKA (a German acronym for Fluctuating Cascade) 2011 \citep{Fas05, Bat07}, GHEISHA (Gamma-Hadron-Electron-Interaction SH(A)ower)2002 \citep{Fes85}, UrQMD (Ultrarelativistic Quantum Molecular Dynamics )\citep{Bas98, Blei99} for hadron interactions below 80 GeV/nucleon and  QGSJET (Quark Gluon String with JETs) II \citep{Ost06} for high energy range above 80 GeV/nucleon. 

COsmic Ray SImulations for KASKADE  (CORSIKA) code is  a widely used
atmospheric cascade simulation tool. The code simulates the
interactions and decays of various nuclei, hadrons, muons,
electrons and photons in the atmosphere. The particles are tracked
through the atmosphere until they undergo reactions with an air
nucleus or in the case of instable secondary particles, they
decay. The result of the simulations is detailed information about
the type, energy, momenta, location and arrival time of the
produced secondary particles at given selected altitude a.s.l. As
primary particles could be considered protons, light and heavy
nuclei up to iron.

\section{Ionization yield function computed with various hadron generators}

The atmospheric cascade simulations are performed with CORSIKA 6.990 code. As was mentioned above, various hadron generators for hadron interactions below 80 GeV/nucleon are assumed: FLUKA 2011 , GHEISHA 2002,  UrQMD.  For high energy range above 80 GeV/nucleon the QGSJET II hadron generator is applied. In addition, various atmospheric models, namely winter, summer and US standard are also applied \citep{Ber00, Kil04, Kil06} in order to study their influence on cosmic ray induced ionization   \citep{Mis08, Mis10a}. During the simulations an isotropic flux of the considered primary CR nuclei is assumed. We simulate up to 100 000 events per energy point per nuclei and compare the ionization yield function $Y$. 

\subsection{Proton nuclei}
The difference of ionization yield function for primary protons computed with various hadron generators is widely discussed in \citep{Mis07b, Mis10a}. Here we include also UrQMD hadron generator. The difference is significant in the case of 1 GeV/nucleon and 10 GeV/nucleon, while above 100 GeV/nucleon the ionization capacity is in practice model independent (Fig.01). In all cases FLUKA produce more ion pares compared to GHEISHA and UrQMD, specifically in the Pfotzer maximum.

\begin{figure}[H]
\begin{center}
\epsfig{file=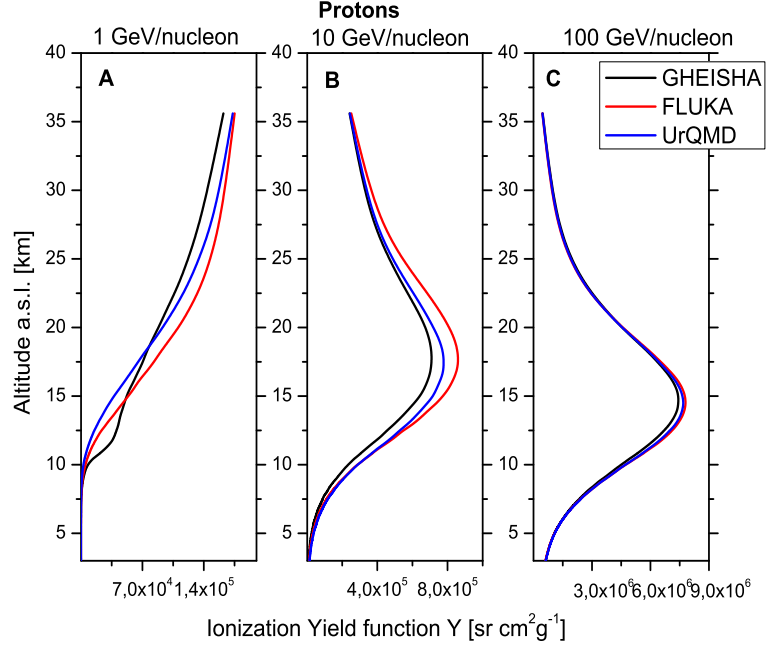,width=11cm, height=9cm} \caption{Ionization yield function for proton nuclei induced atmospheric cascades with energy (a)1 GeV/nucleon, (b) 10 GeV/mucleon, (c) 100 GeV/nucleon assuming FLUKA 2011, GHEISHA 2002 and UrQMD hadron generators.}
\end{center}
\end{figure}

\subsection{$\alpha$-particles}
The simulation results for primary $\alpha$-particles are presented on Figs. 2a-2d for 1 GeV, 10 GeV, 100 GeV and 1 TeV kinetic energy of the primary nuclei. In the case of 1 GeV/nucleon the relative difference between yield functions computed with different hadron generators is irregular as a function of the altitude. A significant difference above about 16 km a.s.l. with excess of ionization capacity assuming FLUKA hadron generator is seen (Fig. 2a). The relative difference between FLUKA and GHEISHA is in the order of 15-18 $\%$ at altitude of 15 km a.s.l., while below and above this level it increases significantly up to 50-80 $\%$ or even to 120-125 $\%$ (at altitude about 11 km a.s.l.). Accordingly, the difference between FLUKA and UrQMD is about $30\%$ in the region in and above the Pfotzer maximum (a secondary particle intensity maximum at the altitude of 15-26 km, which depends on latitude and solar activity)(Fig. 2a). Below this level the relative difference is quasi-constant in the order of $20\%$. The relative difference between UrQMD and GHEISHA is significant in the region of the Pfotzer maximum and diminish below this altitude. In the troposphere at altitude of about 5 km a.s.l., where the intrinsic cascade fluctuations takes over, the relative difference between the hadron generators demonstrate irregular behaviour and could be as high as 40-50 $\%$ (Fig.3).

\begin{figure}[H]
\begin{center}
\epsfig{file=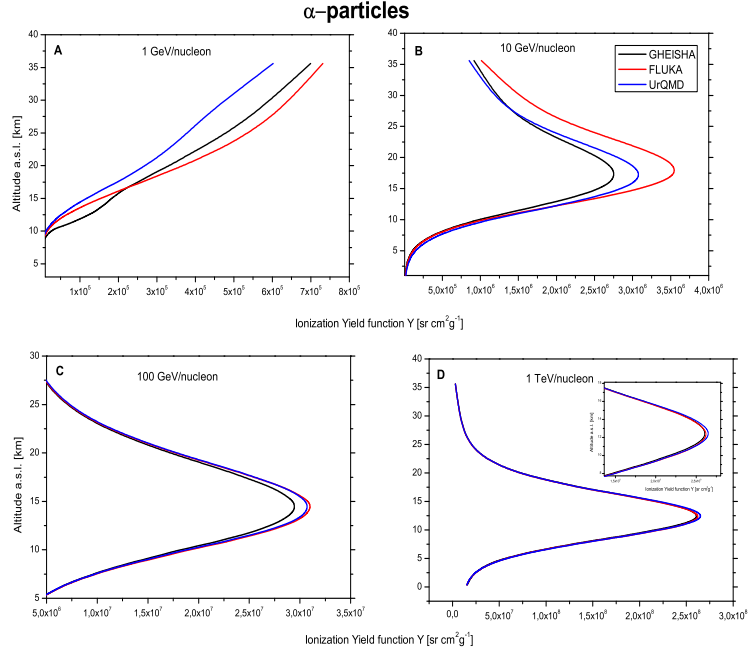,width=11cm,height=10cm} \caption{Ionization yield function for $\alpha$-particles induced atmospheric cascades with energy (a) 1 GeV/nucleon, (b) 10 GeV/mucleon, (c) 100 GeV/nucleon, (d) 1 TeV/nucleon (e) zoom of Pfotzer maximum in the case of 1 TeV/nucleon assuming FLUKA 2011, GHEISHA 2002 and UrQMD.}
\end{center}
\end{figure}

A significant difference for 10 GeV/nucleon $\alpha$-particles(Fig. 2b) is also observed. An excess of ionization capacity assuming FLUKA hadron generator is clearly seen. The difference between FLUKA and GHEISHA is about 15-24 $\%$ at altitude above 14-15 km a.s.l. Below this altitude the difference diminish to 10 $\%$. At altitude below 8 km a.s.l. it drops to the intrinsic cascade fluctuations.  The difference between FLUKA and UrQMD as well between UrQMD and GHEISHA is smaller, specifically in the region of the Pfotzer maximum. At altitude between 10-14 km a.s.l. the relative difference between FLUKA and UrQMD is not significant. In the region of 10-20 km a.s.l. the relative difference between UrQMD and GHEISHA is roughly 10 $\%$, while below this altitude increases to about 20-25 $\%$, which is mostly due to large intrinsic cascade fluctuations. The relative difference is presented in Fig.3.

In the case of 100 GeV/nucleon induced by $\alpha$-particles atmospheric cascades, the difference between FLUKA and GHEISHA is 5-8 $\%$ in the region of the Pfotzer maximum (Fig.2c). The difference above about 20 km a.s.l. and below 10 km a.s.l. is negligible. The difference between FLUKA and UrQMD is in the order of the intrinsic fluctuations of the cascade development (Fig. 2c, Fig.3c). Therefore, the computed ionization capacity is in practice model independent outside the Pfotzer maximum. In the case of 1 TeV/nucleon induced by $\alpha$-particles cascades, we observe difference only in the region of the Pfotzer maximum (Fig.2d) and the zoom of the Pfotzer maximum (Fig.2e). In this case the ionization capacity assuming UrQMD hadron generator is slightly greater compared to other models. The difference is in the order of the intrinsic cascade fluctuations. Therefore, similarly to the case of 100 GeV/nucleon, the ionization yield function is model independent.   

\begin{figure}[H]
\begin{center}
\epsfig{file=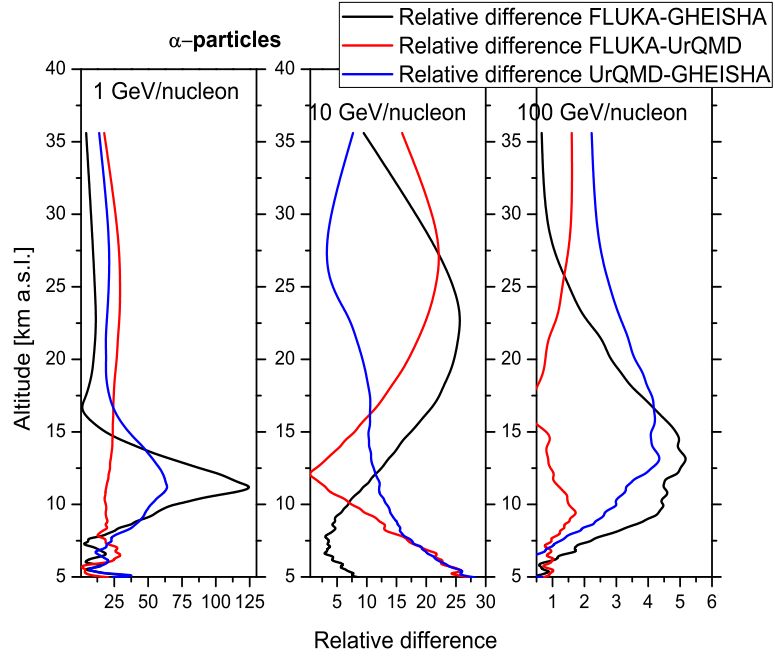,width=11cm, height=9cm} \caption{Relative difference for ionization yield function for $\alpha$-particles assuming FLUKA 2011, GHEISHA 2002 and UrQMD, (a) 1 GeV/nucleon, (b) 10 GeV/nucleon, (c) 100 GeV/nucleon.}
\end{center}
\end{figure}

\subsection{Oxygen nuclei}
The simulation results for primary Oxygen nuclei are presented on Figs. 4a-4d for 1 GeV, 10 GeV, 100 GeV and 1 TeV kinetic energy of the primary. In the case of 1 GeV/nucleon Oxygen nuclei the relative difference between different models is irregular as a function of the altitude. The difference is significant, specifically between FLUKA and UrQMD hadron generators (Fig. 4a). The difference between FLUKA and GHEISHA is quasi-constant above the Pfotzer maximum, with excess of ionization capacity assuming FLUKA hadron generator. Between 12 and 18 km a.s.l. an excess of ionization capacity assuming GHEISHA hadron generator is seen. At altitude of about 10 km a.s.l. the difference between the assumed hadron generators is still significant. Below 10 km a.s.l the difference between  FLUKA and GHEISHA is quasi-constant in the order of 10-15 $\%$, while between FLUKA and UrQMD  increases significantly to 25-30 $\%$. At altitudes below 5 km a.s.l.,  the relative difference between different hadron generators is irregular and could be as high as 70-80 $\%$. An illustration of relative difference is shown in a Fig. 5a. The difference at low altitude is due mostly on a larger cascade to cascade fluctuation and the less number of secondary particles compared to upper altitudes. 

\begin{figure}[H]
\begin{center}
\epsfig{file=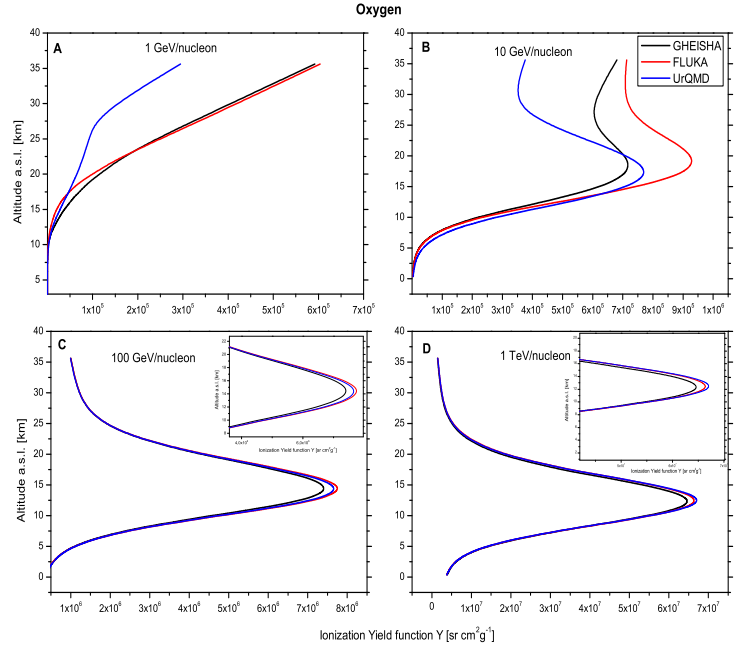,width=11cm,height=10cm} \caption{Ionization yield function for Oxygen nuclei induced atmospheric cascades with energy (a) 1 GeV/nucleon, (b) 10 GeV/mucleon, (c) 100 GeV/nucleon, (d) 1 TeV/nucleon (e) zoom of Pfotzer maximum in the case of 1 TeV/nucleon assuming FLUKA 2011, GHEISHA 2002 and UrQMD.}
\end{center}
\end{figure}

In the case of 10 GeV/nucleon Oxygen nuclei induced cascade a significant difference in the region of the Pfotzer maximum and above (Fig. 4b) is observed. In the region of the Pfotzer maximum an excess of ionization capacity assuming FLUKA hadron generator is clearly seen. The ionization capacity assuming UrQMD hadron model is between FLUKA and GHEISHA, while above this altitude GHEISHA is between FLUKA and UrQMD (Fig. 4b). The difference between FLUKA and GHEISHA in the region of about 10-20 km a.s.l. is 5-25 $\%$, while between FLUKA and UrQMD drops to about  1-3 $\%$ at altitude of 14 km a.s.l. The difference varies between 15 to 20 $\%$ in the other cases. At altitude below 10 km a.s.l. the difference between FLUKA and GHEISHA is quasi-constant in the order of 5 $\%$, while between FLUKA and UrQMD is larger: $\approx$ 20 $\%$. The relative difference is presented in Fig. 5b.

In the case of 100 GeV/nucleon Oxygen induced atmospheric cascade (Fig. 4c) the difference between FLUKA and GHEISHA is greater than between FLUKA and UrQMD. The relative difference between FLUKA and GHEISHA is slightly above the intrinsic shower fluctuations, while between FLUKA and UrQMD is in the order of statistical and intrinsic cascade fluctuations (Fig.5c). In this case a slight excess of ionization capacity assuming FLUKA hadron generator is seen. Similar situation is observed in the case of 1 TeV/nucleon Oxygen induced atmospheric cascade (Fig. 4d). We conclude: the ionization yield function $Y$ for Oxygen nuclei is in practice model independent in the energy region above 100 GeV/nucleon. 

\begin{figure}[H]
\begin{center}
\epsfig{file=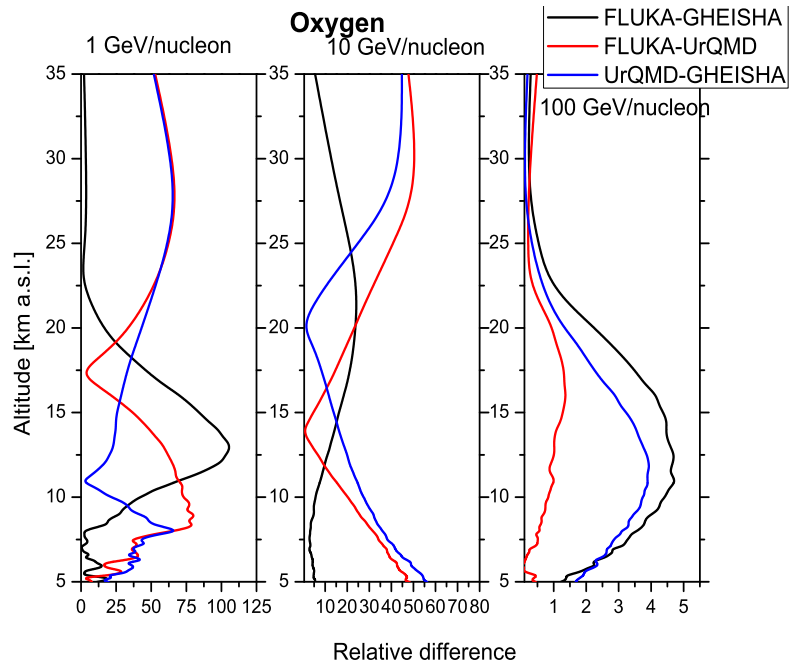,width=11cm, height=9cm} \caption{Relative difference for ionization yield function for Oxygen nuclei assuming FLUKA 2011, GHEISHA 2002 and UrQMD, (a) 1 GeV/nucleon, (b) 10 GeV/nucleon, (c) 100 GeV/nucleon.}
\end{center}
\end{figure}

\subsection{Iron nuclei}
Similar comparison is performed for Iron nuclei induced atmospheric cascades (Figs. 6a-b). In the case of 1 GeV/nucleon (Fig. 6a) an important difference is observed between FLUKA and UrQMD, accordingly between GHEISHA and UrQMD. The difference between FLUKA and GHEISHA is 55-95 $\%$ in the region 10-20 km a.s.l. and rapidly diminish at altitudes below 10 km a.s.l. The difference between FLUKA and UrQMD is significant, specifically in the region of the Pfotzer maximum. The relative difference is presented in Fig. 7a. 

\begin{figure}[H]
\begin{center}
\epsfig{file=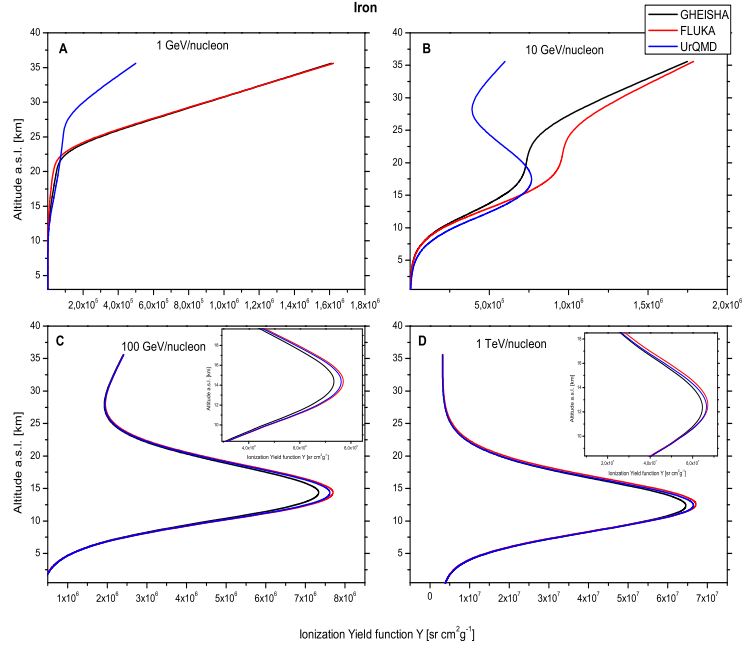,width=11cm,height=10cm} \caption{Ionization yield function for Iron nuclei induced atmospheric cascades with energy (a) 1 GeV/nucleon, (b) 10 GeV/mucleon, (c) 100 GeV/nucleon, (d) 1 TeV/nucleon (e) zoom of Pfotzer maximum in the case of 1 TeV/nucleon assuming FLUKA 2011, GHEISHA 2002 and UrQMD.}
\end{center}
\end{figure}

In the case of 10 GeV/nucleon Iron induced cascades (Fig. 6b) a complicated behaviour of ionization capacity in a whole atmosphere is observed. The difference between FLUKA and GHEISHA is roughly 15 $\%$ at altitude above 25 km a.s.l., while in the region of the Pfotzer maximum the difference is about 20 $\%$. The difference between FLUKA and GHEISHA below 10 km a.s.l. is about 5-8 $\%$ (see Table 1). The difference between FLUKA and UrQMD is larger at altitude above 20 km a.s.l. $\approx$ 30-40 $\%$, while in the region of the Pfotzer maximum it drops to about 10-15 $\%$ (at altitude of 15 km a.s.l. is roughly 2 $\%$).

In the case of 100 GeV/nucleon and 1 TeV/nucleon Iron induced atmospheric cascades (Fig. 6c and Fig. 6d) the situation is very similar to the case for Oxygen nuclei. In both cases a difference of about 5-8 $\%$ between FLUKA and GHEISHA, specifically in the Pfotzer maximum  is observed. The difference between FLUKA and UrQMD is smaller. It is in the order of the intrinsic cascade fluctuations. Therefore, we can conclude: the ionization yield function $Y$ for Iron nuclei is model independent in the energy region above 100 GeV/nucleon.

\begin{figure}[H]
\begin{center}
\epsfig{file=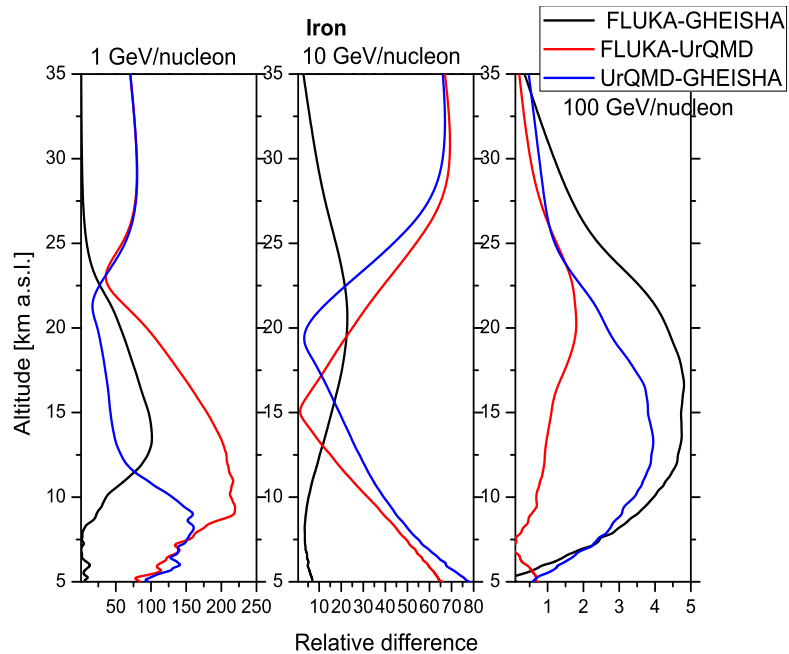,width=11cm, height=9cm} \caption{Relative difference for ionization yield function for Iron nuclei assuming FLUKA 2011, GHEISHA 2002 and UrQMD, (a) 1 GeV/nucleon, (b) 10 GeV/nucleon, (c) 100 GeV/nucleon.}
\end{center}
\end{figure}

The discussed above effects are due mainly to the influence of various hadron interaction models on atmospheric cascade  development, which leads to different energy deposit, accordingly ionization yield function (2). The influence of hadron generators on ionization yield function for proton induced cascades is widely discussed in \citep{Mis07b, Mis10a} and for nuclei in \citep{Mis13a, Mis14}. Furthermore, as was previously reported small changes in a medium (atmosphere) properties could lead to changes of ionization capacity \citep{Mis08, Mis10a}. Therefore, assuming various atmospheric models, namely US standard atmosphere, US standard atmosphere winter and summer profiles, the influence of atmosphere parametrization on cosmic ray induced ionization is estimated see below.

\section{Seasonal effect: Applications and Discussion}
It the work presented here, the applied low energy hadron generators are: the very refined model with many details of nuclear effects FLUKA, a microscopic model used to simulate (ultra)relativistic heavy ion collisions (UrQMD) and the well approved Monte Carlo code for detector response simulation (GHEISHA). The influence of the model assumptions of various hadron interaction models on atmospheric cascade simulation, the difference and comparison of hadron generators for atmospheric cascade simulation, their influence on cascade development and energy release are widely discussed in several papers (see \citet{Hec03, Risse04, Hec06, Engel11} and references therein). In addition, as it was demonstrated for primary CR protons and nuclei, the different atmospheric parametrizations lead to approximatively 10-15 $\%$ difference of ion production rate. The detailed comparison of assumed hadron generators as well as the seasonal effect due to atmospheric profile variation at various rigidity cut-off is given below. 

The contribution of protons and $\alpha$-particles to the total ionization due to CR is roughly 50 $\%$, specifically in the upper atmosphere. The contribution of heavier CR nuclei (M, H and VH groups) is approximatively the same, despite their lower abundance. A significant difference (Figs 1-7) of ionization capacity, specifically in the region of Pfotzer maximum is seen. Therefore, a detailed comparison of ion rate production assuming various models is necessary. Hence, we perform a detailed Monte Carlo simulation of atmospheric cascade considering realistic CR mass composition \citep{Nak10} and force field  model of galactic CR (GCR) spectrum  \citep{Gle68, Cab04, Mc04, Usos06}, using the approximation of local interstellar spectrum \citep{Uso05}. In this case no normalization of ionization yield function is performed \citep{Usos06, Mis09, Mis11a, Mis11b} since the ionization yield function per nucleon $Y/nucleon$ for the different CR ions differs significantly, specifically in the region above the Pfotzer maximum (Fig.8a-c). This difference is the main motivation for direct simulation of heavy nuclei induced cascades instead of scaling or substitution similarly to \citet{Usos06, Mis11a, Mis11b} and several successful applications \citet{Koval12, Mis13}. 

\begin{figure}[H]
\begin{center}
\epsfig{file=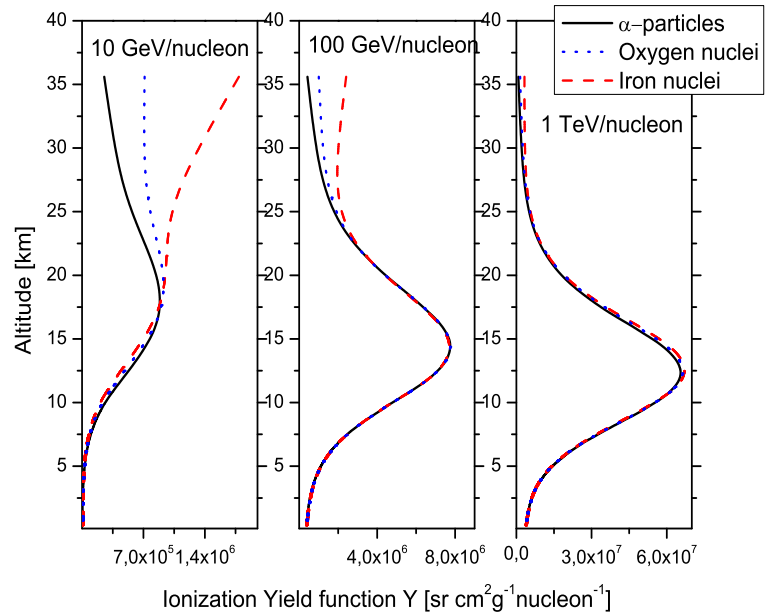,width=11cm, height=9cm} \caption{Ionization yield function per nucleon $Y/nucleon$ for 10 GeV/nucleon (a), 100 GeV/nucleon (b), 1 TeV/nucleon (c) induced atmospheric cascades by various nuclei ($\alpha$-particles, Oxygen and Iron).}
\end{center}
\end{figure}

The computed by us ionization yield functions demonstrate good agreement with previously reported results \citep{Usos06}. As it is seen on Fig. 8, the substitution of heavy nuclei with the corresponding number of $\alpha$-particles \citep{Usos06} is reasonable at the region of the Pfotzer maximum and below (Fig.8a-c). The direct simulation of heavy nuclei in this case lead to similar result in a low atmosphere \citep{Usos06}. However, for precise and realistic estimation of ion production rate in the upper part of the middle atmosphere all primary CR nuclei should be considered separately as it is seen in Fig. 8a-c. This lead to improvement of agreement between models and experimental data in the order of at least 30 $\%$. 

The ion production rates due to GCR (mean solar modulation) assuming US standard atmosphere parametrized by \citet{Kil04} and various hadron generators at 1 GV rigidity cut-off are presented on a left panel of Fig.9. The ion rate is scaled to maximum. At altitudes above 18 km a.s.l. the ion production rate profiles assuming FLUKA and GHEISHA hadron generators are in practice the same. They are greater than the ion production rate due to GCR assuming UrQMD hadron generator. In the region of the Pfotzer maximum ion rate profiles assuming GHEISHA and UrQMD hadron generator are very similar, roughly 85 $\%$ of ion rate produced by FLUKA. Below the Pfotzer maximum, the ion production rate assuming FLUKA is still greater than the produced by GHEISHA/UrQMD. Below this altitude UrQMD slightly produces more ion pares comparing to GHEISHA. Moreover, a slight difference of the position of the Pfotzer maximum is observed, with highest altitude produced assuming GHEISHA hadron generator and lowest level assuming UrQMD hadron generator. On the right panel of Fig.9 we compare the ion production rate produced by FLUKA and GHEISHA assuming various atmospheric parametrizations, namely US standard parametrized by \citet{Kil04}, US standard winter and summer parametrizations \citep{Hec97}. The ion rate estimated with different CORSIKA code model assumptions (hadron generator and atmospheric parametrization) is compared with experimental data \citep{Neh67, Neh71, Baz08}. The measurements of \citet{Neh67, Neh71} are performed during the summer period, namely June-July using modified ionizing chamber \citep{Neher56}. In general, the ion rate assuming winter or summer atmospheric parametrization differs roughly with 10 $\%$. In addition, the US standard atmosphere following the parametrization of \citet{Kil04} is between summer and winter US standard atmospheric parametrization. The obtained with FLUKA and QGSJET II ion rate is the most consistent with the experimental data in the region of the Pfotzer maximum. GHEISHA and UrQMD also demonstrate good agreement with the experiment below the Pfotzer maximum. 

\begin{figure}[H]
\begin{center}
\epsfig{file=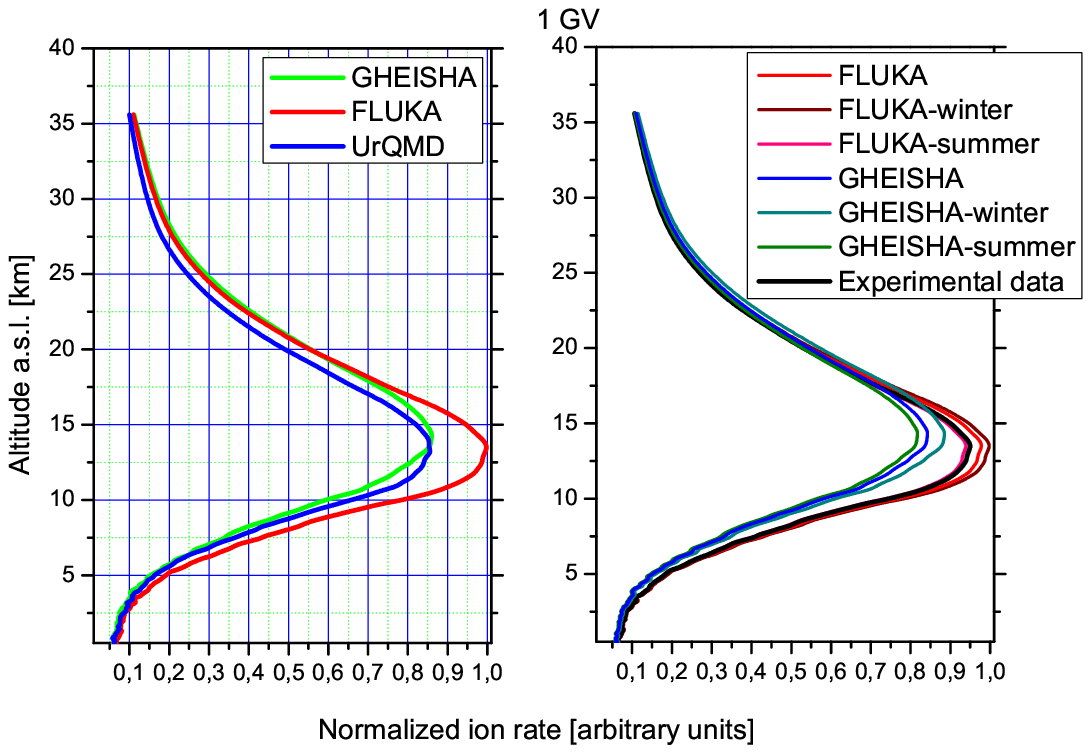,width=11cm, height=9cm} \caption{Scaled to maximum ion production rate at 1 GV rigidity cut-off  (left panel) obtained assuming FLUKA 2011, GHEISHA 2002 and UrQMD hadron generators and US standard atmosphere parametrized according \citet{Kil04}. Right panel: ion production rate produced by FLUKA and GHEISHA assuming US standard atmosphere parametrized by \citep{Kil04}, US standard atmosphere winter and summer parametrizations \citet{Hec97} compared with experimental data based on balloon-born measurements in June-July 1965 \citep{Neh71}.}
\end{center}
\end{figure}

The ion production rate at 5 GV rigidity cut-off assuming US standard atmosphere parametrized by \citet{Kil04} obtained with various hadron generators is presented on the left panel of Fig.10. In the region of the Pfotzer maximum the ion production rate assuming FLUKA hadron generator is greater then the produced by GHEISHA/UrQMD. Above the Pfotzer maximum at altitude about 22-23 km a.s.l. dominates FLUKA. In the region above 23 km a.s.l. the ion rate profiles assuming various hadron generators are the same. In the region below approximatively 8 km a.s.l. slight increase produced by UrQMD hadron generator is observed. In general the difference between FLUKA and GHEISHA is greater then between FLUKA and UrQMD. On the right panel of Fig.10 we present the ion production rate produced by FLUKA and GHEISHA assuming various atmospheric parametrizations, namely US standard parametrized by \citep{Kil04}, US standard winter and summer parametrizations \citet{Hec97} compared with experimental data based on balloon-born measurements in June-July 1965 \citep{Neh67}. A good agreement between FLUKA and experimental data is achieved. 

\begin{figure}[H]
\begin{center}
\epsfig{file=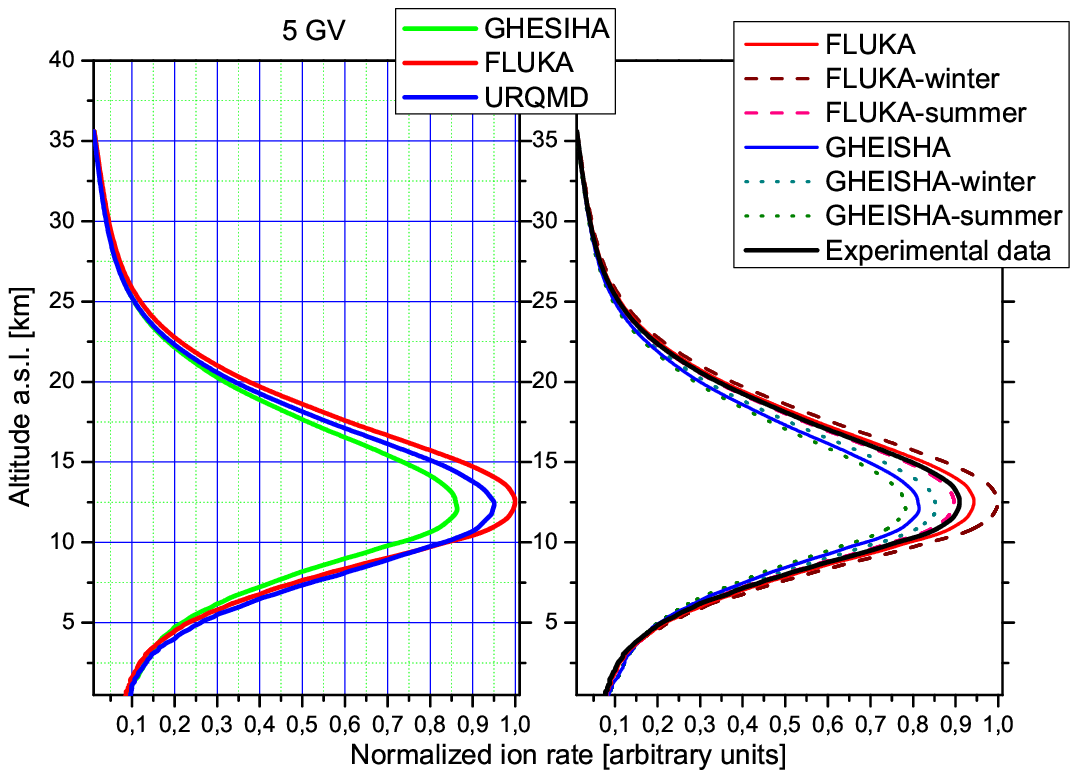,width=11cm, height=9cm} \caption{Scaled to maximum ion production rate at 5 GV rigidity cut-off (left panel) obtained assuming FLUKA 2011, GHEISHA 2002 and UrQMD hadron generators and US standard atmosphere parametrized according \citet{Kil04}. Right panel: ion production rate produced by FLUKA and GHEISHA assuming US standard atmosphere parametrized by \citep{Kil04}, US standard atmosphere winter and summer parametrizations \citet{Hec97} compared with experimental data based on balloon-born measurements in June-July 1965 \citep{Neh67}.}
\end{center}
\end{figure}

The ion production rate profiles at 9 GV and 15 GV are very similar (Fig.11). In both cases we observe difference in the region of the Pfotzer maximum. The ion production rate assuming FLUKA hadron generator is greater then the produced by GHEISHA (roughly 85 $\%$ of ion rate produced by FLUKA) and UrQMD (about 95 $\%$ of ion rate produced by FLUKA). At altitude above 20 km a.s.l. and below 8 km a.s.l. the ion rate is model independent. 

\begin{figure}[H]
\begin{center}
\epsfig{file=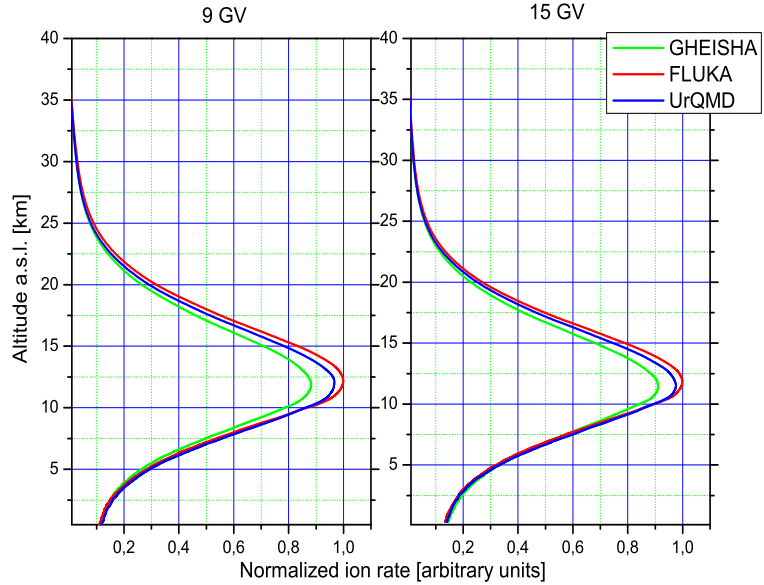,width=11cm, height=9cm} \caption{Scaled to maximum ion production rate at 9 GV rigidity cut-off (left panel) and 15 GV rigidity cut-off (right panel) obtained assuming FLUKA 2011, GHEISHA 2002 and UrQMD hadron generators and US standard atmosphere parametrized according \citet{Kil04}.}
\end{center}
\end{figure}

The derived ion production rate profiles in the atmosphere obtained with different hadron generators assuming various atmospheric parametrizations are not limited only to GCR. As it was recently shown, the ion rate production due to some major solar energetic particle events could be strong locally \citep{Uso11, Mis11, Mis11c, Mis12a, Mis12b, Mis13c}, specifically in sub-polar and polar regions, affecting the physical-chemical properties of the upper atmosphere \citep{Kri06, Rep10}. Therefore, the detailed knowledge of model constraints, respectively simulation tool are very important for adequate interpretation of observations and models. Obviously further studies as well as development and improvement of analytical model for cosmic ray induced ionization, specifically in region above the Pfotzer maximum are necessary \citep{Vel13a, Vel13b}.   

\section {Summary}
The effect of model assumptions on Monte Carlo simulations of cosmic ray induced ionization are important, because they are related to explanation and realistic modelling of different processes in the atmosphere. In the work presented here, we demonstrate the influence of different hadron generators, namely GHEISHA 2002, FLUKA 2011, UrQMD on cosmic ray induced ionization. The study focus on the contribution of heavy CR nuclei, which is important for precise modelling of GCR induced ionization as well as the ion rate production due to major solar energetic particles events. The ion production rate profiles  are obtained at various rigidity cut-offs (1GV, 5 GV, 9GV and 15 GV) assuming realistic primary CR mass composition and mean GCR flux. The difference is observed essentially in the region of the Pfotzer maximum. In general the obtained ion rate profiles are with similar shape. Below the Pfotzer maximum are in practice the same.  The applied in this study low energy hadron interaction models are fully applicable at low altitude, however it seems that FLUKA (Figs.9-11) is the most consistent with experimental data and previous results \citep{Hec03, Hec06}.

A seasonal difference of ion production rate in the region of the Pfotzer maximum of about 10-15 $\%$ is observed  assuming different atmospheric parametrizations (US standard atmosphere winter, respectively summer profile).  

The presented results are important for studies related with influence of cosmic rays on atmospheric processes, space weather (for details see \citet{Vel13a} and references therein),could contribute significantly to recent studies of cosmic ray induced ionization \citep{Usos06, Baz08, Uso09, Mishev12, Mis13b, Mis13c} and  are applicable in several domains of solar-terrestrial physics \citep{Mir03}.

\textbf{\Large Acknowledgments}\\
We acknowledge the high-energy division of Institute for Nuclear Research and Nuclear Energy - Bulgarian Academy of Sciences for the given computational time. We warmly acknowledge Prof. Ilya Usoskin from Oulu University-Finland for the discussions related to the application of cosmic ray induced ionization model.


\end{document}